\documentclass[lettersize,journal]{IEEEtran}
\usepackage{amsmath,amsfonts}
\usepackage{algorithmic}
\usepackage{algorithm}
\usepackage{array}
\usepackage[caption=false,font=normalsize,labelfont=sf,textfont=sf]{subfig}
\usepackage{textcomp}
\usepackage{stfloats}
\usepackage{url}
\usepackage{verbatim}
\usepackage{graphicx}
\usepackage{cite}
\usepackage{tikz}   
\usepackage[hidelinks]{hyperref} 
\usepackage{orcidlink}   

\usepackage[normalem]{ulem}
\usepackage{xcolor}
\usepackage{color}

\begin{document}

\title{Optimization procedure of the baffle of the GroundBIRD Telescope to mitigate stray light}

\author{
Miku Tsujii\,\orcidlink{0009-0000-6741-6033},
Tomonaga Tanaka,
Alessandro Fasano\,\orcidlink{0000-0003-4041-418X},
Ricardo Génova-Santos\,\orcidlink{0000-0001-5479-0034},
Shunsuke Honda\,\orcidlink{0000-0002-0403-3729},
Yonggil Jo\,\orcidlink{0000-0002-4340-3171},
Keisuke Kataoka,
Chiko Otani\,\orcidlink{0000-0002-9406-2602},~\IEEEmembership{Member,~IEEE},
Mike Peel\,\orcidlink{0000-0003-3412-2586},
Junya Suzuki\,\orcidlink{0000-0001-6816-8123},
Osamu Tajima\,\orcidlink{0000-0003-2439-2611},
Eunil Won\,\orcidlink{0000-0002-4245-7442},
and Makoto Hattori\,\orcidlink{0000-0003-0620-2554}
\thanks{\textcopyright~2026 IEEE. Personal use of this material is permitted.
Permission from IEEE must be obtained for all other uses, in any current or future media,
including reprinting/republishing this material for advertising or promotional purposes,
creating new collective works, for resale or redistribution to servers or lists, or reuse
of any copyrighted component of this work in other works.
Published in IEEE Transactions on Applied Superconductivity.
DOI: \href{https://doi.org/10.1109/TASC.2026.3672265}{10.1109/TASC.2026.3672265}.
(Corresponding author: Miku Tsujii.)}
\thanks{M. Tsujii, T. Tanaka, and M. Hattori are with the Astronomical Institute, Tohoku University, Sendai, Miyagi 980-8578, Japan  (e-mail: miku.tsujii@astr.tohoku.ac.jp; hattori@astr.tohoku.ac.jp)}
\thanks{A. Fasano and R. Génova-Santos are with the Instituto de Astrofísica 
de Canarias (IAC), E-38200 La Laguna, Tenerife, Spain, and also with the 
Departamento de Astrofísica, Universidad de La Laguna (ULL), E-38206 La Laguna, 
Tenerife, Spain}
\thanks{S. Honda is with the Division of Physics, Faculty of Pure and Applied Sciences, University of Tsukuba, Tsukuba, Ibaraki 305-8571, Japan, and also with Tomonaga Center for the History of the Universe (TCHoU), University of Tsukuba, Tsukuba, Ibaraki 305-8571, Japan}
\thanks{Y. Jo and E. Won are with the Department of Physics, Korea University, 
Seoul 02841, South Korea}
\thanks{K. Kataoka, J. Suzuki, and O. Tajima are with the Department of Physics, 
Faculty of Science, Kyoto University, Kyoto, Kyoto 606-8502, Japan}
\thanks{C. Otani is with the RIKEN Center for Advanced Photonics, RIKEN, 
Sendai, Miyagi 980-0845, Japan, and also with the Department of Physics, 
Tohoku University, Sendai, Miyagi 980-8578, Japan}
\thanks{M. Peel is with Imperial College London, London SW7 2AZ, UK}
}
 
\maketitle

\begin{abstract}
We present the optimization procedure of the baffle mounted on the GroundBIRD telescope for measuring the polarization of the Cosmic Microwave Background~(CMB). The telescope employs a dual mirror reflective telescope installed in a cryostat.
The primary objectives were to minimize stray light contamination, maintain the integrity of the main beam, and ensure that thermal loading from the baffle remains significantly below that from the atmosphere.
Using quasi-optical simulations, we have optimized the baffle's aperture angle to suppress stray light without degrading the main beam quality.
We confirmed through Moon observations that the optimized baffle design works to eliminate the contamination of the stray light as expected. Furthermore, no measurable degradation in the noise equivalent temperature~(NET) was detected, indicating minimal thermal impact. 
These results show that our baffle optimization strategy effectively reduces systematic errors while maintaining observational sensitivity, providing valuable insights for future CMB experiments with similar optical architectures.
\end{abstract}


\section{Introduction}
\begin{figure}[!t]
\centering
\includegraphics[width=3.in]{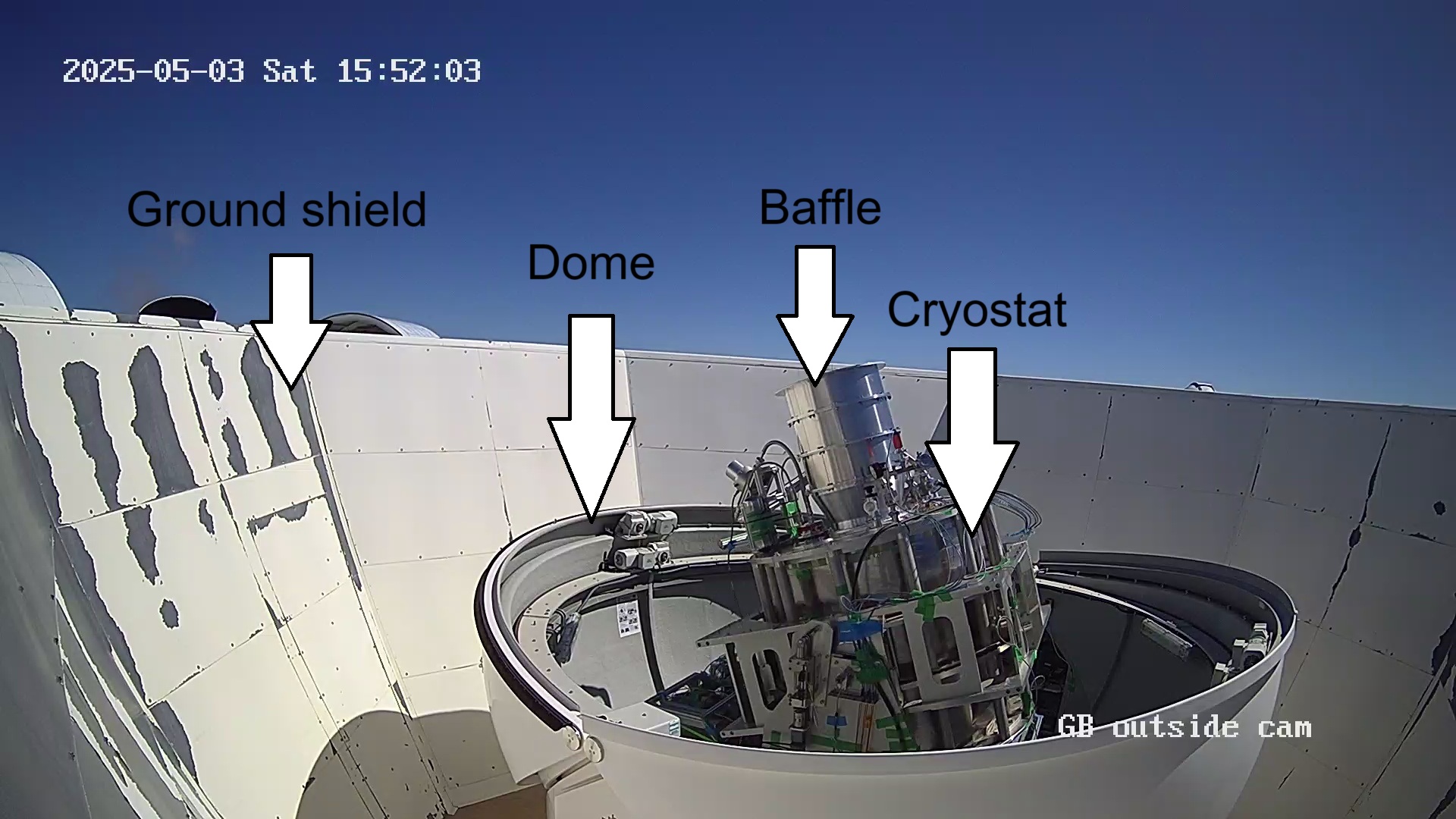}
\caption{A photograph of the GroundBIRD telescope installation captured by a monitoring camera. The arrangement of the ground shield, dome, baffle, and cryostat is shown.}
\label{fig:GB_image}
\end{figure}

\IEEEPARstart{T}{he} precision measurements of the Cosmic Microwave Background~(CMB) 
observations have been playing a central role in exploring the space-time geometry, evolution, and origin of our universe\cite{dodelson:2003, komatsu:2019}. 
The current challenge of CMB observations is high-precision measurements of its polarization signals. 
One of the key issues in CMB experiments is reducing systematic errors in the data to achieve the required high accuracy of the measurements. 
Suppressing the penetration of stray light into the detectors is one of the key subjects for reducing
systematics. 
To prevent contamination of light reflected by the cryostat wall into the detectors, all walls of the cryostat are covered with high-performance millimeter wave absorbers\cite{adachi2020production, xu2021simons, otsuka2021material}.
In ground-based CMB experiments, either a fixed ground shield\cite{takahashi2010characterization, galitzki2024simons} or a co-moving ground shield\cite{Kusaka_2018, harrington2016cosmology, arnold2010polarbear, austermann2012sptpol} is applied.
In some CMB experiments, a baffle installed in front of the entrance window plays a complementary role in reducing the contamination of the ground signal\cite{takahashi2010characterization, galitzki2024simons, matsumura2014mission}. 
Fig.~\ref{fig:GB_image} shows the installation of the GroundBIRD telescope.

\begin{figure}[!t]
\centering
\includegraphics[width=2.3in]{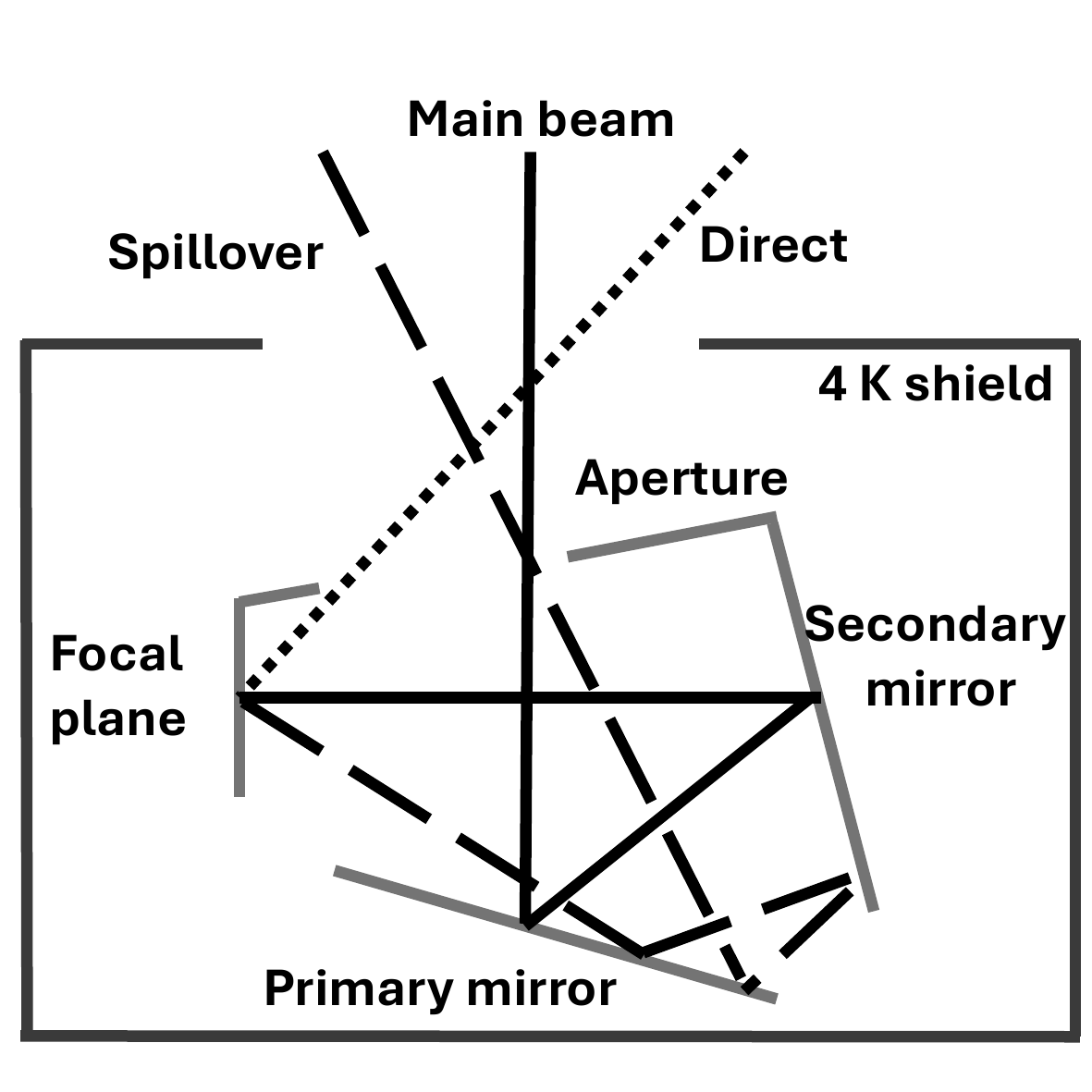}
\caption{Primary stray light paths in the GroundBIRD optical system, including key optical components, are illustrated. The main beam is drawn by a solid line. 
Two types of stray light paths are illustrated: the ``spillover" component~(dashed line) and the ``direct" component~(dotted line).}
\label{fig:LightPass}
\end{figure}

In the case of telescopes adopting reflective mirrors\cite{austermann2012sptpol, matsumura2014mission, genova2015quijote, dicker2018cold, fowler2007optical, jo2023simulation}, to avoid diffraction from optical components and their mounting structures in the light path, an off-axis reflecting mirror system is employed. 
To compensate for the instrumental polarization imprinted by the off-axis reflection, the secondary mirror is positioned to satisfy the Mizuguchi-Dragone conditions, which minimize contamination of instrumental polarization\cite{dragone1978offset}.
Two mirror configurations are proposed to satisfy the Mizuguchi-Dragone conditions: the crossed configuration and the Gregorian configuration.
The crossed configuration has advantages for realizing a wider field of view and less cross-polarization compared to the Gregorian configuration\cite{Tran:08}. 
However, 
the crossed configuration is more susceptible to scattering than the Gregorian configuration\cite{Tran:08}.
The Gregorian configuration, such as SPT\cite{austermann2012sptpol} and ACT\cite{fowler2007optical}, has a real intermediary focus between the secondary and primary mirrors, which can be exploited to effectively shield the focal plane from direct illumination, making it less susceptible to far sidelobes\cite{Tran:08}. In contrast, the crossed configuration adopted by GroundBIRD\cite{jo2023simulation} presents a more difficult baffling situation because the focal plane is close to the main beams incident on the primary mirror\cite{Tran:08}, resulting in higher susceptibility to far sidelobes due to scattering from structures near the focal plane.
As shown by \cite{jo2023simulation}, two stray light components reach the detectors in the GroundBIRD telescope~(see Fig. \ref{fig:LightPass}). 
One is reaching the detectors without experiencing any reflection by the primary and secondary mirrors.
It is referred to as the direct component of the stray light.
The other is reflected again by the primary mirror after reflection by the secondary mirror before reaching the detectors.
It is referred to as the spillover component of the stray light. 
One way to reduce stray light contamination on the detectors is to increase the distance from the secondary mirror to the detectors by using lenses enclosed in the lens barrel\cite{dicker2018cold}.
The lens barrel serves as an absorber, blocking stray light. 
However, in some experiments\cite{jo2023simulation, sekimoto2020concept, matsuda2024sidelobe, Matsuda:25}, it is difficult to increase the distance
between the secondary mirror and the detectors due to space limitations.
In these cases, a baffle plays a role in preventing penetration of these stray light components into the detectors. 
To suppress the contamination of the light reflected by the wall of the baffle, the wall of the baffle is also covered by millimeter wave absorbers. 

There are two ways to reduce 
contamination from stray light by optimizing the design of the baffle.
One is extending the baffle length. 
The other is narrowing the top aperture diameter of the baffle. 
In the case of the GroundBIRD experiment, 
the baffle length was set to the 
maximum value allowed by the dome size~(see Fig.~\ref{fig:GB_image}).
Therefore, narrowing the top aperture diameter of the baffle is the only method for reducing contamination from stray light.
However, narrowing the top aperture diameter of the baffle is accompanied by the risk that the main beams are degraded due to diffraction from the baffle aperture.
In this study, we present the optimization procedure 
for the top baffle aperture diameter to reduce stray light contamination
while maintaining the quality of the main beams. 
We show how the quality of the main beams and stray light rejection depend on the top aperture diameter of the baffle of the GroundBIRD based on the results of the simulations and Moon observations. 
Although performing wave optics simulations is ideal for the optimization, 
we adopt the quasi-optical simulation instead
because the physical optics simulation is computationally expensive.

The structure of the paper is as follows. 
In Section II, we first introduce the instrumental details of GroundBIRD and its observation strategy.  
We also introduce the simulation setup. 
In Section III,  the methodology for optimizing the baffle design is presented. 
Section IV presents the details of the optimized  baffle design. 
Section V presents the results of the performance verification measurements. 
Section VI is devoted to discussion and conclusion. 

\section{Methods}
\subsection{GroundBIRD Telescope\label{subsec:GroundBIRD}}
The GroundBIRD telescope is a ground-based CMB polarization experiment designed to capture the reionization bump signals imprinted in the large angular scale of approximately 20 degrees in the northern hemisphere for the first time from the ground after the Planck satellite\cite{collaboration2020planck, essinger2014class}.
It is considered that it is challenging to reach the reionization bump from ground-based observations due to
atmospheric fluctuations, although the CLASS experiment has conducted observations in the southern hemisphere to contribute to large-scale polarization measurements\cite{Li_2025}.
GroundBIRD is attempting to reach the reionization bump through ground-based observations by mitigating atmospheric fluctuations
with its unique scan strategy\cite{tajima2012groundbird}. 
It performs a rapid rotation scan of up to 20 rotations per minute~(RPM) while inclining the telescope 20 to 30 degrees from the zenith. 
By scanning large patches of the sky before the sky condition changes significantly, it has the potential to capture the reionization bump. 

\begin{figure}[!t]
\centering
\includegraphics[width=2.in]{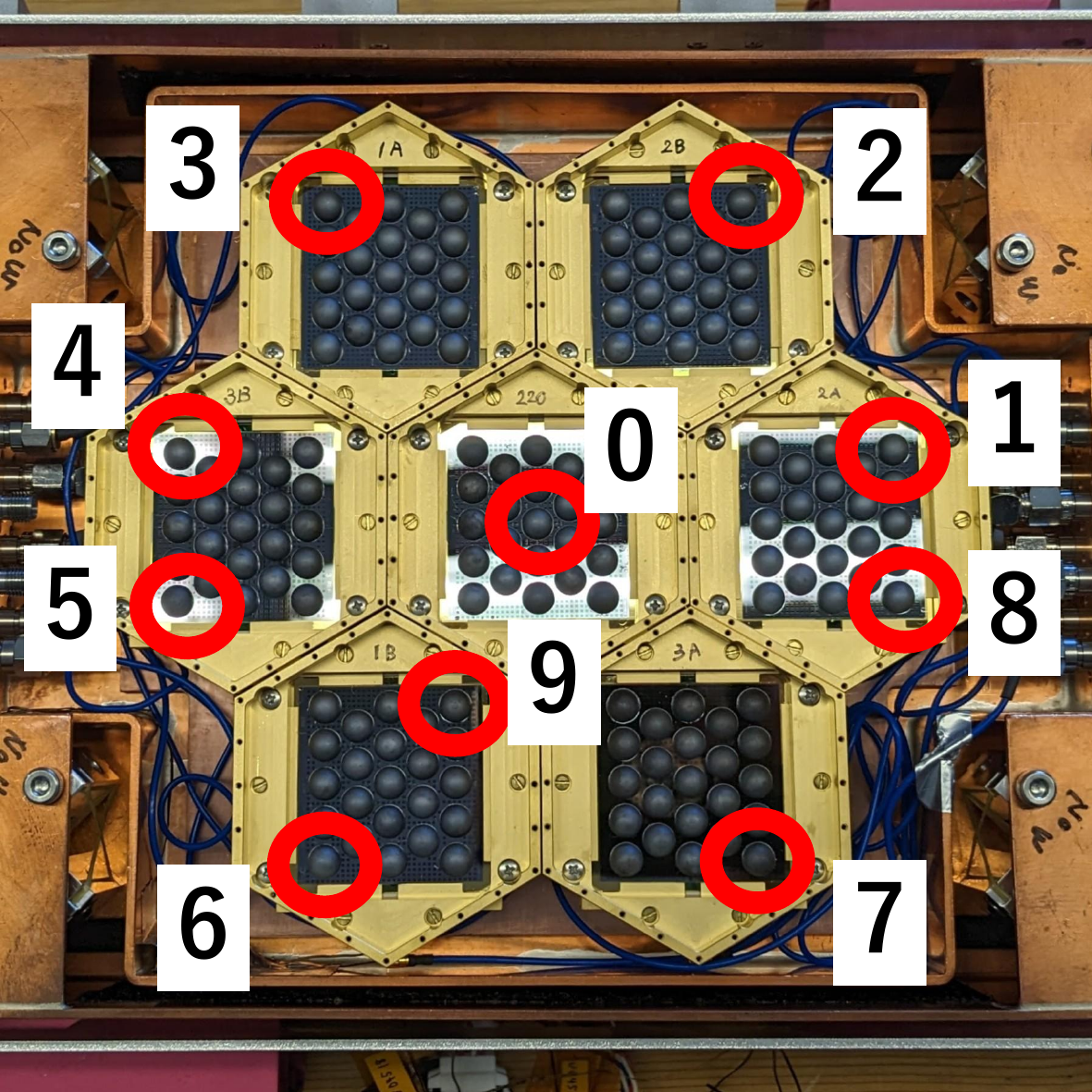}
\caption{A photograph of the detector array mounted on the focal plane. The pixels for which we show the results of our analysis are assigned numbers. Pixels numbered from 1 through 8 are selected because their thermal and optical characteristics are most affected by baffle design.}
\label{fig:pixelIDs}
\end{figure}

To achieve diffraction-limited spatial resolution simultaneously, cutting-edge superconducting detectors, Microwave Kinetic Inductance Detectors~(MKIDs), are utilized as the focal plane detectors due to their fast time response.
The observed frequency bands are 145 GHz and 220 GHz. 
Assembly of the detectors is shown in Fig.~\ref{fig:pixelIDs}. 
The central chip mounts 23 pixels for 220 GHz band and the other 6 chips mount 138 pixels for 145 GHz band.
GroundBIRD employs Mizuguchi-Dragone dual reflectors\cite{dragone1978offset}.
The whole telescope system is installed in the cryostat
and cooled down to 4 K. 

The telescope is set at the Teide observatory in Tenerife, Canary Islands.
First light was in 2019\cite{honda2020first}, and the full detector set was installed in 2023\cite{tsujii2024commissioning}.

\subsection{Quasi-Optical Simulations With GRASP\label{subsec:Simlation setup}}
We quantified near-field and far-field beam patterns for each pixel using quasi-optical simulations.
Although we ultimately want to determine how each detector pixel observes the sky—that is, the beam pattern for received light—we instead calculate the beam pattern emitted from each detector pixel. This is justified by the time-reversal invariance of Maxwell's equations, which ensures that both patterns are identical.

\begin{figure*}[!t]
\centering
\subfloat[]{\includegraphics[width=2.2in]{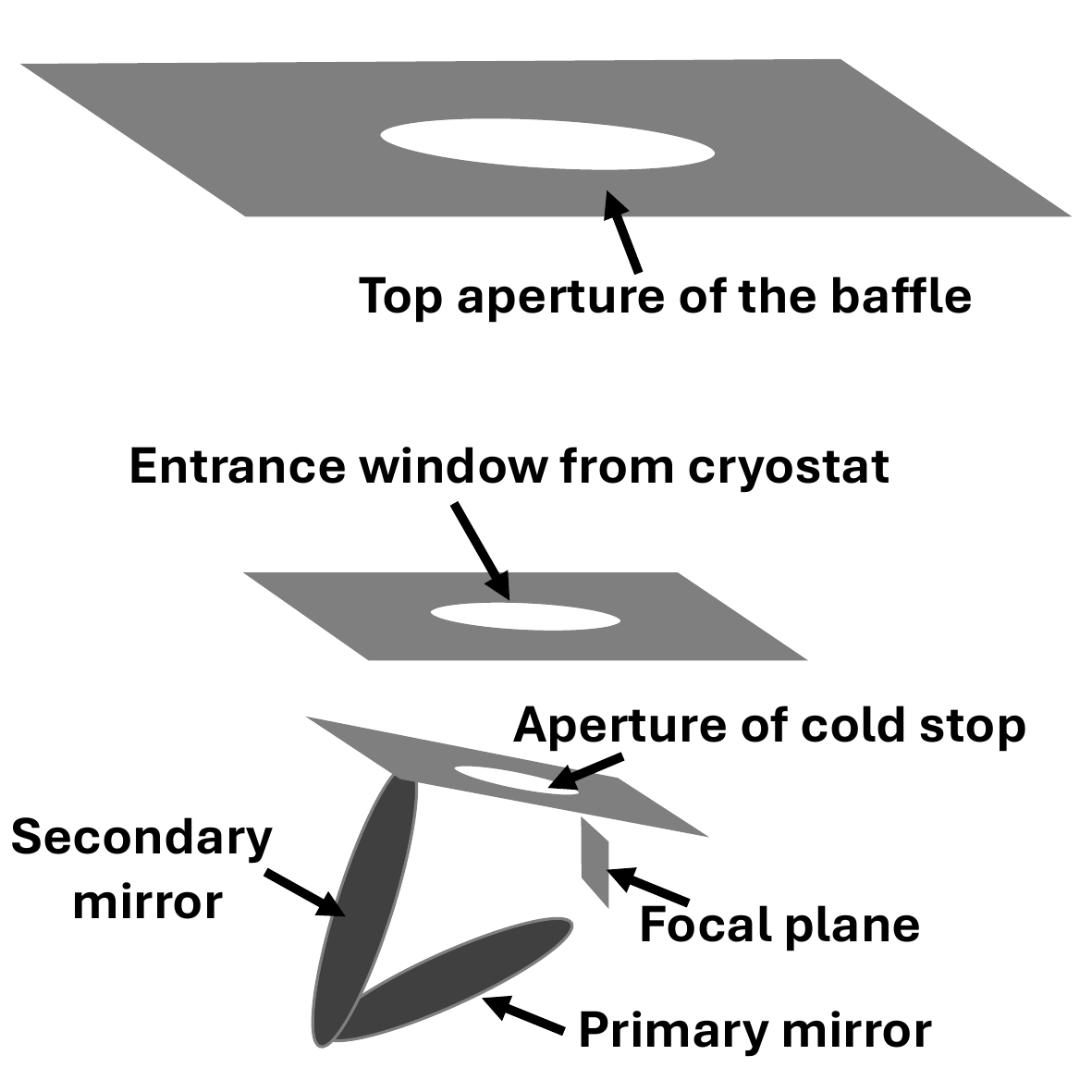}
\label{fig:GRASP_far}}
\hfil
\subfloat[]{\includegraphics[width=1.8in]{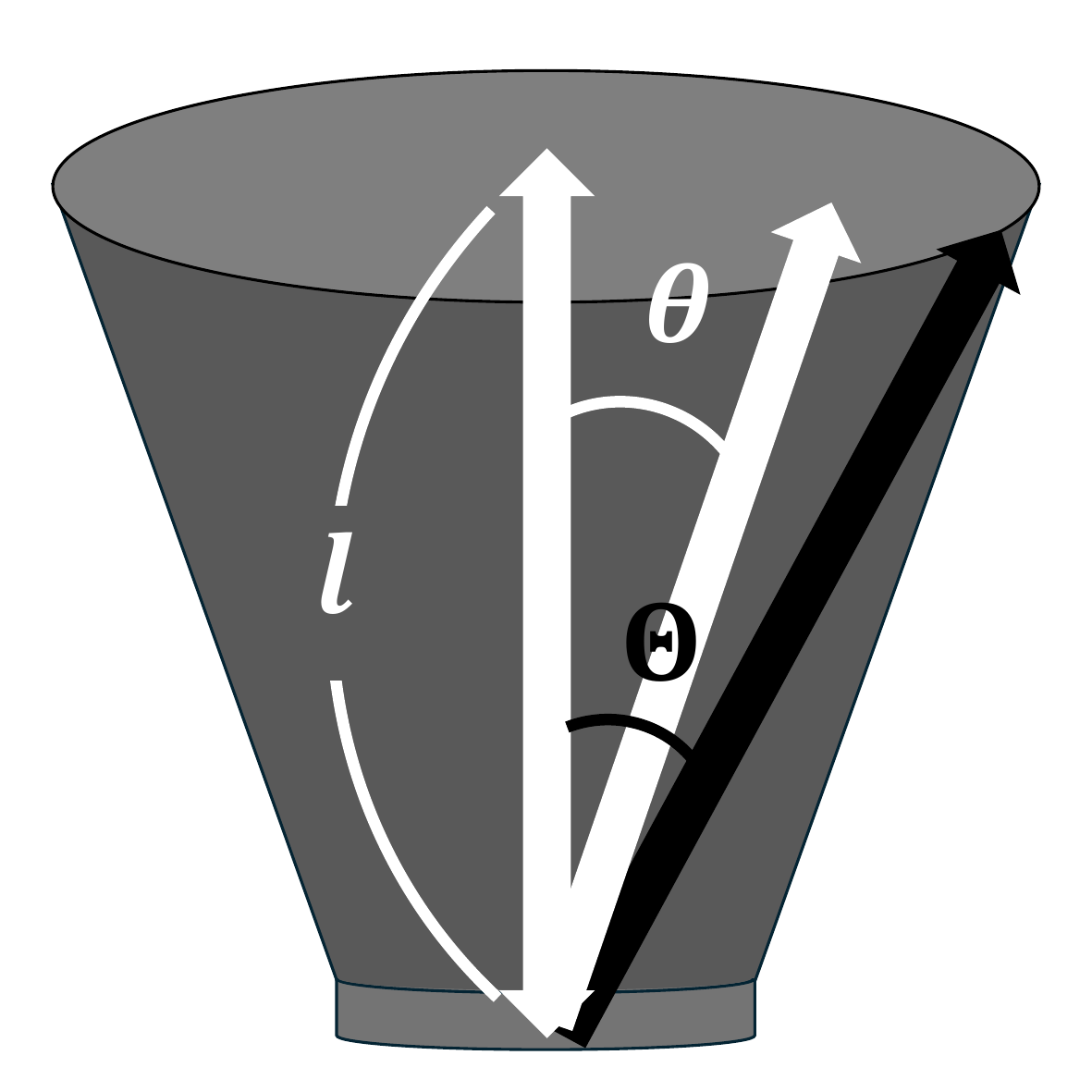}
\label{fig:baffle}}
\caption{(a) The simulation setup for the quasi-optical simulations with GRASP. The white holes in the gray areas represent the optical components. The top circular area represents the entrance aperture at the top of the baffle. The entrance window from the cryostat and the aperture of the cold stop, which defines the entrance pupil, are shown by 
another two circular areas. 
(b) The definition of the parameters that define the principal geometry of the baffle.
}
\end{figure*}

The antenna attached to each MKID first emits the beam.
Then, it is modified by the lenslet mounted on each MKID. 
Up to this stage, the beam pattern is calculated by using a physical optics simulation with CST\cite{cst}. 
Beam patterns after the lenslet are quantified by quasi-optical simulations with GRASP\cite{grasp}. 
The results of CST are used for boundary conditions for the GRASP simulations.
GRASP simulations are performed for both far-field and near-field beam patterns. 
The setup for far-field calculations is shown in Fig.~\ref{fig:GRASP_far}, where all optical components including the top aperture of the baffle are incorporated. 
For near-field calculations, the top aperture of the baffle is excluded to evaluate thermal loading from the baffle wall without diffraction effects.
The apertures of the baffle, the entrance window of the cryostat, and the cold stop are set. 
The detector pixel position is set according to the design.
To set the positions of the primary and the secondary mirrors, the CAD model is used. 
The mirrors are modeled as perfect conductors. 
The aperture of the cold stop, which defines the entrance pupil, has a circular shape with a diameter of 220 $\mathrm{mm}$. 
The entrance window from the cryostat also has a circular shape with a diameter of 320 $\mathrm{mm}$.
The plane of the cold stop is tilted relative to the entrance window plane to avoid standing waves due to the Fabry-Perot effect.
Parameters describing the characteristics of the baffle are shown in Fig.~\ref{fig:baffle}. 
To quantify the top aperture diameter of the baffle, the opening angle $2\Theta$ is used.
The aperture angle, $\Theta$, is defined by the arctangent of the aperture radius divided by the length of the baffle.
The baffle length is the distance between the top of the entrance window of the cryostat and the top aperture of the baffle. 
In our optimization procedure, the baffle length was set to the 
maximum value~(82 cm) allowed by the dome, and only the aperture angle $\Theta$ was varied to suppress stray light while maintaining beam quality.
This is the physical length of the baffle plus 10 mm. 
The angular distance relative to the optical axis of the dual mirror system is denoted by $\theta$. 

Validations of the simulated beam patterns were performed using Moon observations. 
Although Jupiter and Venus are observable bright point sources, they are too faint to characterize the beam patterns down to the side lobes within a single-day measurement.
Thanks to the wide dynamic range of the MKIDs, it is possible to observe the Moon with GroundBIRD.
The Moon is sufficiently bright for characterizing the beam pattern down to the side lobes including stray light with one-hour integration time.
Since the Moon has to be treated as an extended source by GroundBIRD, 
the specific power observed during Moon observations by detector pixel ID $i$ at frequency $\nu$ is given by 
\begin{equation}
    P_{\nu,i}(\alpha_i, \beta_i) = \frac{1}{2} A_e\ t_r(\nu)\int_{4\pi} d\Omega \ U_{\nu,i}(\alpha-\alpha_i, \beta-\beta_i) I_{\nu}(\alpha, \beta)\label{eq:conv}
\end{equation}
where $\alpha$ and $\beta$ are spherical polar coordinates of a unit sphere centered on the Moon where the origin coincides with the Moon center, $(\alpha_i, \beta_i)$ is a direction of the beam center of the i-th detector relative to the Moon center, the factor $\frac{1}{2}$ accounts for single polarization detection, $A_e$ is the effective aperture area, $t_r(\nu)$ is the transmittance of the telescope's filter including the quantum efficiency of the detector, $U_i(\alpha-\alpha_i, \beta-\beta_i)$ is the beam pattern of detector pixel $i$, and $I_{\nu}(\alpha, \beta)$ is the intensity distribution of the Moon at frequency $\nu$.
We modeled the Moon's intensity as a blackbody sphere covered by a dielectric material layer \cite{bischoff2010phd} and adapted the model described in \cite{sueno2024pointing}.

\section{Baffle Design Optimization}
\subsection{Beam Quantification With a Wide Aperture Angle Baffle}
\begin{figure}[!t]
\centering
\subfloat[]{\includegraphics[width=0.45\linewidth]{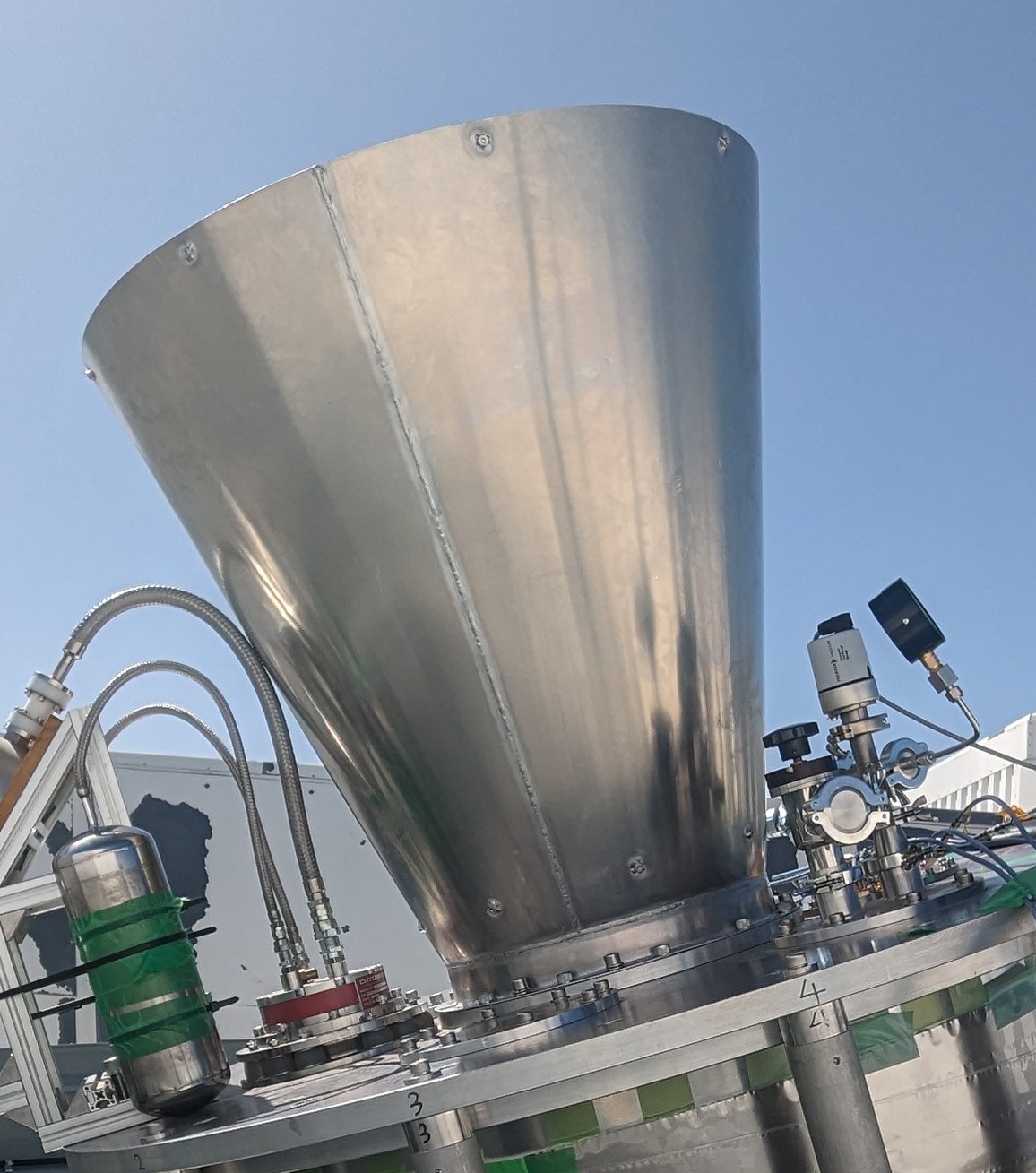}
\label{fig:OldBaffle_out}}
\hfill
\subfloat[]{\includegraphics[width=0.5\linewidth]{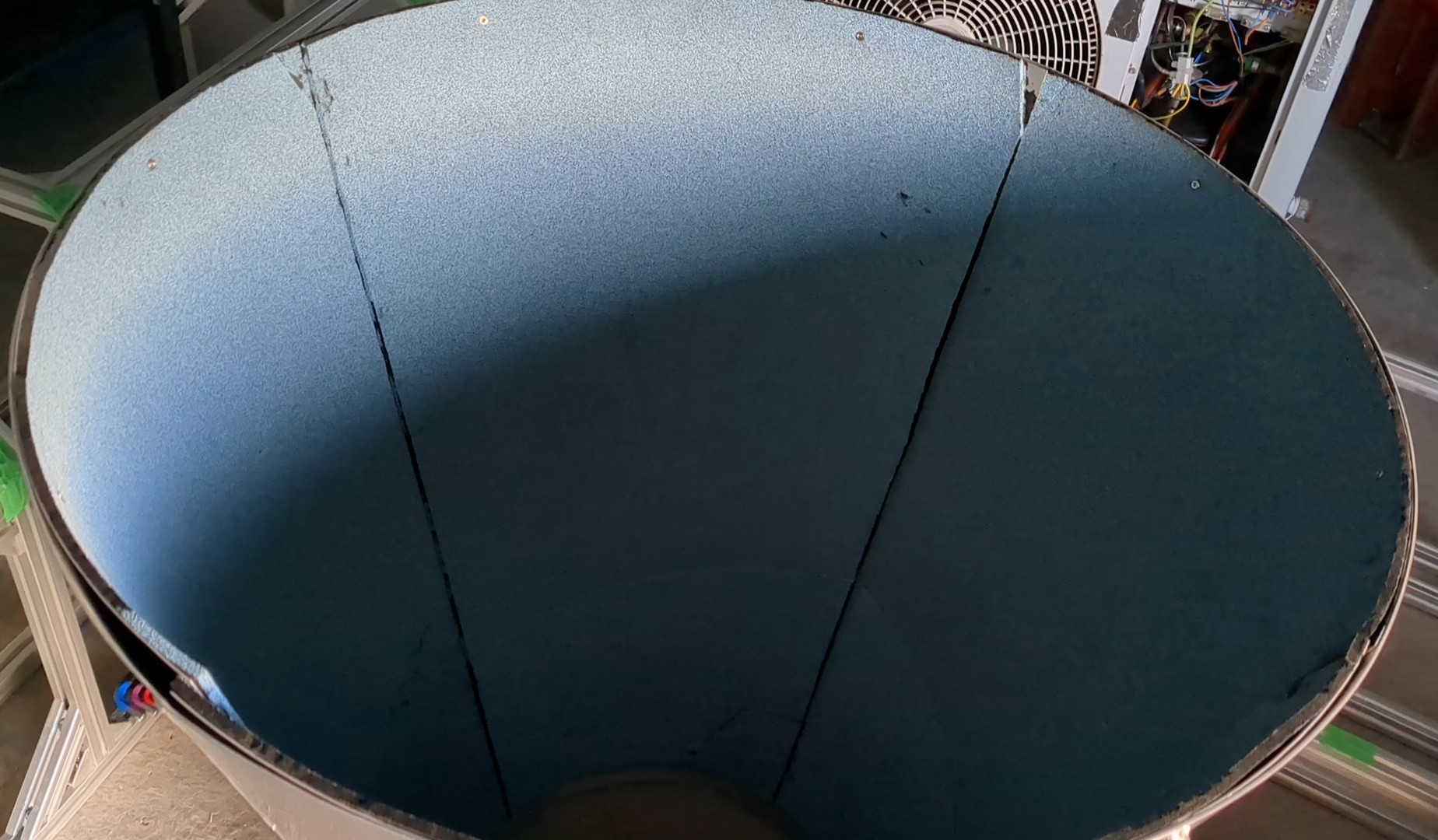}
\label{fig:OldBaffle_in}}
\caption{A photograph of the wide baffle mounted on the GroundBIRD.
The baffle length is 84 cm and the aperture angle $\Theta =27^{\circ}$.
(a) External view. (b) Internal view showing the radio-absorptive material~(ECCOSORB AN-72) covering the inner wall. The absorber is bonded to the aluminum surfaces using Stycast 2850FTJ adhesive.}
\label{fig:original_baffle}
\end{figure}
\begin{figure*}[!t]
\centering
\subfloat[]{\includegraphics[width=2.8in]{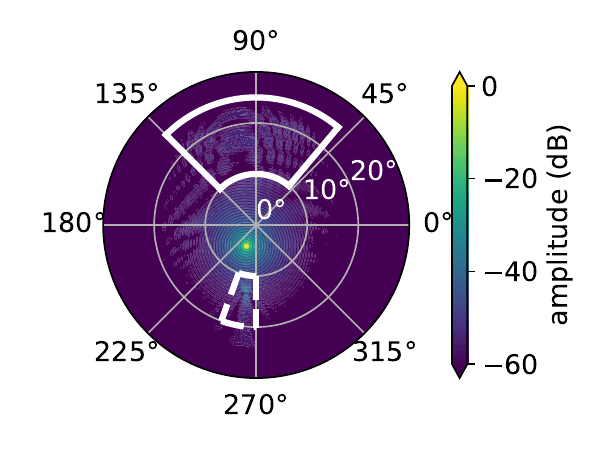}
\label{fig:82cm_9}}
\hfil
\subfloat[]{\includegraphics[width=2.8in]{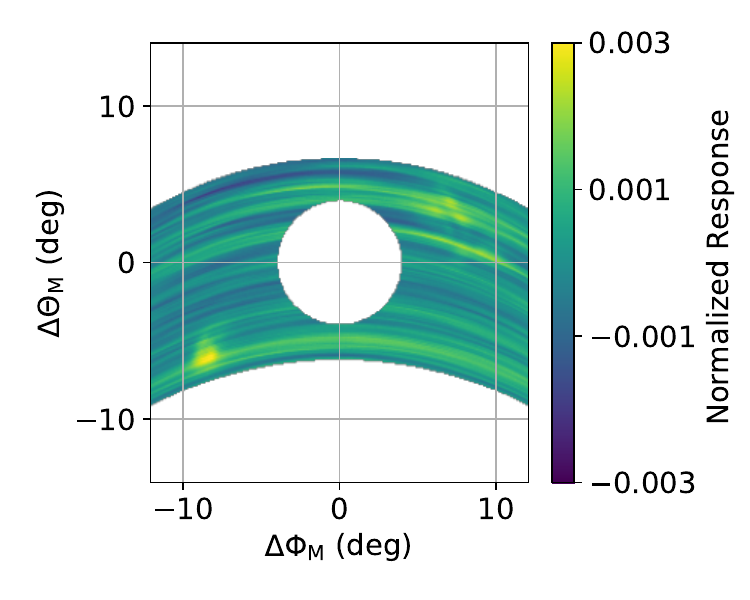}
\label{fig:moon_stray_original}}
\caption{(a) Simulated beam pattern for the pixel ID 9 with the wide baffle~($\Theta = 27^\circ$). The relative angular distance from the pointing center, $\theta$, is drawn by concentric circles with 10 degrees interval. The azimuth angles, $\phi$, are drawn by radial lines with 45 degrees intervals. 
The region where the spillover component appears is enclosed by the white dashed line. 
The region where the direct component appears is enclosed by the white solid line.
(b) Observed map of the Moon obtained by the pixel ID 9 when the wide baffle is mounted on the GroundBIRD. Vertical and horizontal axes are the offset angle relative to the Moon center in the elevation, $\Delta \theta_\mathrm{M}$, and azimuth, $\Delta \phi_\mathrm{M}$, directions respectively.  
The maps are masked within a 4-degree radius from the Moon's center and divided into grids with 0.1-degree intervals. For each scan, baseline noise was subtracted by fitting a linear function to $\pm$1500 data points from $\Delta \phi_\mathrm{M}=0$, while masking the central $\pm$500 data points. The intensity is normalized to the maximum intensity of the Moon image, where the detector response is proportional to $\tan(\frac{\psi}{2})$ and $\psi$ is a phase of fed microwaves for the detector readout\cite{Gao2008}.}
\end{figure*}

The beam patterns at $\nu = 145$ GHz when using the baffle with the wide aperture angle corresponding to $\Theta = 27^{\circ}$~(see Fig.~\ref{fig:OldBaffle_out}) were simulated. 
In this study, we refer to the baffle with $\Theta = 27^{\circ}$ as the ``wide baffle" for convenience.
Fig.~\ref{fig:82cm_9} shows the simulated far field 2D beam pattern for the detector pixel denoted by ID 9 as shown in Fig.~\ref{fig:pixelIDs}.
The dashed line encloses the region where the spillover component appears.
We can see some localization of the beam in the central region enclosed by the dashed line.
On the other hand, 
the direct component appears sparsely in the region enclosed by the solid line which is much wider than the region enclosed by the dashed line without any prominent structure. 
The peak intensity of the direct component is about two orders of magnitude less than the peak intensity of the spillover component. 
Since the spillover component appears closer to the pointing center compared to the direct component, 
it is expected that the direct component is suppressed automatically when the baffle is designed so as to suppress the spillover component.
The specific power when the Moon is picked up by the beam center, $P_{\nu,i,\mathrm{main}}$, is calculated 
by setting $(\alpha_i,\beta_i)=(0,0)$ in~(\ref{eq:conv}). 
The specific power when the Moon is picked up by the spillover component, $P_{\nu,i ,\mathrm{stray}}$, is calculated by setting $(\alpha_i,\beta_i)$ in~(\ref{eq:conv})
so that the specific power of the Moon picked up by the spillover component becomes maximum.  
This yields $P_{\nu,i,\mathrm{stray}}/P_{\nu,i,\mathrm{moon}}\sim-34$ dB for pixel ID 9. 

Fig.~\ref{fig:moon_stray_original} shows a Moon-centered map obtained with the detector pixel ID 9 with the wide baffle. 
The data were acquired during observations on December 2, 2023. 
The observation began at 02:52 UTC and lasted for one hour, encompassing one complete observation cycle between the two successive calibration measurements of the resonance frequencies of the MKIDs.
The average value of the precipitable water vapor~(PWV) was around 2 mm. 
The inclination angle of the telescope from the zenith was 20 degrees.
The telescope was rotated at 10 RPM. 
The Moon image obtained by the main beam is masked to emphasize the ghost image picked up by the stray light. 
We observed a bright spot in the bottom left of the map. 
The relative position of the bright spot relative to the Moon's center is consistent with the relative position of the peak of the spillover component to the peak of the main beam shown in Fig.~\ref{fig:82cm_9}.
Moon observations were conducted multiple times, and we analyzed six observation datasets spanning from December 2, 2023, to October 25, 2024.
The sky conditions were similar in these observations: the PWV was around 1.7 mm -- 2.5 mm. We confirmed that the bright spot appears at the same location in the maps of all observations. Therefore, we concluded that the bright spot observed in Fig.~\ref{fig:moon_stray_original} is the ghost image of the Moon picked up by the spillover component of the stray light. Its peak amplitude compared to that of the Moon center is approximately $-25$ dB.
The background noise level is approximately $-26$ dB.
Therefore, the net peak amplitude of the ghost image after subtracting the background noise is approximately $-30$ dB.
The observed intensity of the ghost image is a few times stronger than that from the beam pattern simulation. 

\subsection{Baffle Aperture Angle Size Dependence Of the Thermal Loading From the Baffle}\label{subsec:thermalload}
\begin{figure}[!t]
\centering
 \includegraphics[width=2.6in]{3/beam\_82cm\_spherical\_5.pdf}
\caption{The cross section of the simulated near-field beam pattern for the pixel ID 5.}
\label{fig:edge_near}
\end{figure}
In this section, the baffle aperture angle dependence of the thermal loading from the absorbers put on the baffle was examined. 
The inner wall of the baffle is entirely covered by radio-absorptive material~(ECCOSORB AN-72)~\cite{eccosorb_an72} bonded to the inner surfaces using Stycast 2850FTJ adhesive~(Henkel Corporation)~\cite{loctite_stycast}~(see Fig.~\ref{fig:OldBaffle_in}).
The absorber emits blackbody radiation corresponding to the ambient temperature of approximately 293 K.
Therefore, the thermal emission from the baffle can be one of the external thermal loads. 
Since the location of the baffle is well within the Fraunhofer distance of $R = 2D^2/\lambda$, that is about 45 m for 145 GHz and 70 m for 220 GHz, the near-field beam pattern must be evaluated to estimate the thermal loading from the baffle. 

\begin{table}[t!]
\caption{The simulated powers carried into each pixel from the atmosphere and baffles.\label{tab:therm_load}}
\centering
\begin{tabular}{|c|c|c|c|c|c|}
\hline
Pixel & Freq. & $P_{\mathrm{atm}}$ & \multicolumn{3}{c|}{$P_{\mathrm{baffle}}$ ($10^{-12}$ W)}  \\
ID & (GHz) & ($10^{-12}$ W) & $\Theta=15^\circ$ & $\Theta=18^\circ$ & $\Theta=20^\circ$\\ 
\hline
        0  & 220 & 6.6 & 0.08 & 0.07 & 0.07\\
        1  & 145 & 3.9 & 2.2 & 0.27 & 0.07\\
        2  & 145 & 3.6 & 2.8& 0.53 & 0.09\\
        3  & 145 & 3.6 & 2.7 & 0.52 & 0.09\\
        4  & 145 & 3.9 & 2.1 & 0.26 & 0.07\\
        5  & 145 & 4.0 & 1.9 & 0.23 & 0.06\\
        6  & 145 & 4.0 & 2.8 & 0.44 & 0.08\\
        7  & 145 & 4.0 & 2.7 & 0.43 & 0.08\\
        8  & 145 & 4.0 & 1.9 & 0.24 & 0.06\\
        \hline
\end{tabular}
\end{table}
To estimate the thermal loading from the baffle, we approximated the GroundBIRD telescope as being enclosed by an open top hemisphere.
Assuming blackbody radiation at 293 K from this sphere, the thermal loading from the baffle is calculated as:
\begin{multline}
    P_{\mathrm{baffle}} = \frac{1}{2}A_{\rm lens} \int d\nu  \int_0^{2\pi} d\phi \int_{\Theta}^{\pi/2} \sin{\theta} d\theta \\
    \times B(\nu, 293\,\mathrm{K})\ t_r({\nu})\ U_{\mathrm{near}}(\theta, \phi)
\end{multline}
where $A_{\rm lens}=\pi (3 {\rm mm})^2$ is the area of the cross section of the Silicon lenslet. 
The integration by $\theta$ is performed from $\Theta$ to $\pi/2$.
The factor $\frac{1}{2}$ accounts for single polarization detection and $B(\nu, 293~{\mathrm{K}})$ represents the Planck function at $293~{\mathrm{K}}$.
We estimate the thermal loading due to the atmospheric radiation by
\begin{equation}
    P_{\mathrm{atm}} =  \frac{1}{2}\eta_{\mathrm{ap}} \int d\nu (1-e^{-\tau_{\nu}}) t_r(\nu) \lambda^2 B(\nu, T=293 {\mathrm{K}})
\end{equation}
where $\tau_{\nu}$
is the optical depth of the atmosphere at frequency $\nu$ and $\eta_{\mathrm{ap}}$ is the aperture efficiency calculated by GRASP. 
Since most of the atmospheric emission originates from the far field and its intensity distribution is nearly uniform across the field of view of each detector pixel, we assume that the effective area times the field of view equals $\lambda^2$.
The thermal loading from the baffle and the atmospheric emission are compared in Table~\ref{tab:therm_load} for the detector pixels labeled in Fig.~\ref{fig:pixelIDs}. 
We assumed that GroundBIRD was pointing 20 degrees from the zenith and atmospheric PWV is 1 mm which is the optimal value representing good weather conditions at the Teide observatory.
It shows that if we take $\Theta=15^{\circ}$, the thermal loading of the baffle is comparable to the atmospheric contribution. 
However, as far as $\Theta$ is larger than 18 degrees, the thermal loading due to the baffle is less than 15~\% of the atmospheric contribution. 
Fig.~\ref{fig:edge_near} shows the cross section of the near field beam pattern for pixel ID 5 crossing its peak power position and $\theta=0$ as an example of the pixels located at the edge of the focal plane.
It shows that a significant portion of the main beam of this pixel overlaps with the baffle when $\Theta=15^\circ$ is selected. 
Since the power of the main beam starts to decrease drastically at $\theta \sim - 15^\circ$, the thermal loading from the baffle also decreases drastically when the aperture angle of the baffle becomes wider than $\Theta > 15^\circ$.  
It shows that at $\theta = - 20^\circ$, the power of the near field main beam is as small as $-20$ dB relative to its peak power.
\subsection{Baffle Aperture Angle Dependence of 
the Far Field Beam Pattern}\label{subsec:faredge}
We checked impacts on the beam pattern with respect to the baffle aperture angle.
\begin{figure*}[!t]
\centering
\subfloat[]{\includegraphics[width=2.6in]{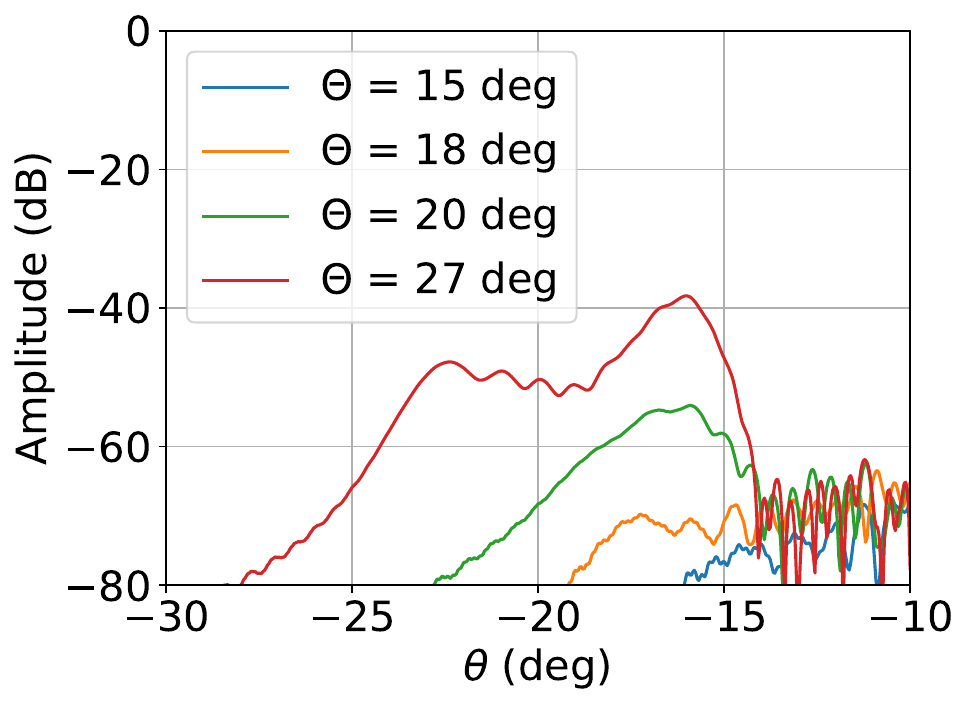}
\label{fig:crosec_spill_0}}
\hfil
\subfloat[]{\includegraphics[width=2.6in]{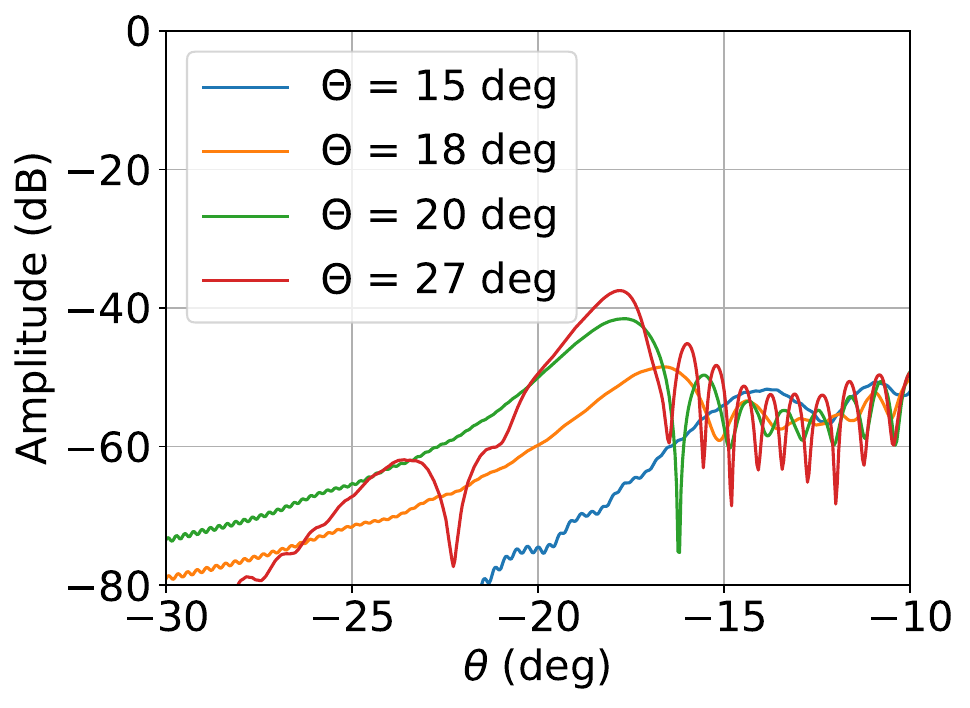}
\label{fig:crosec_spill_5}}
\caption{The cross section of simulated beam with $\Theta=15^\circ, 18^\circ, 20^\circ$ and the wide baffle~($\Theta = 27^\circ$).
The amplitudes are the relative power to the peak power in each map.
These figures demonstrate that reducing the aperture angle effectively suppresses the spillover amplitude, with smaller angles yielding significantly lower spillover contamination.
(a) The center pixel 0.
(b) The edge pixel 5.\label{fig:crosec_spill}}
\end{figure*}
\begin{figure*}[!t]
\centering
\subfloat[]{\includegraphics[width=2.4in]{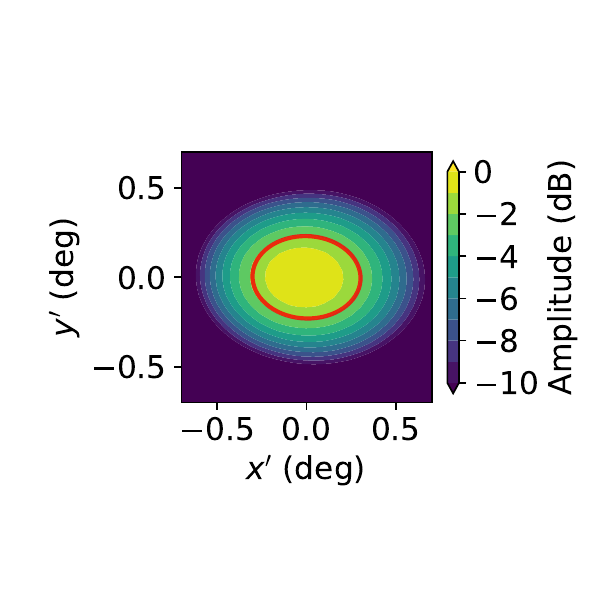}
\label{fig:beam2d_pixel5_15}}
\hspace{-0.2in} 
\subfloat[]{\includegraphics[width=2.4in]{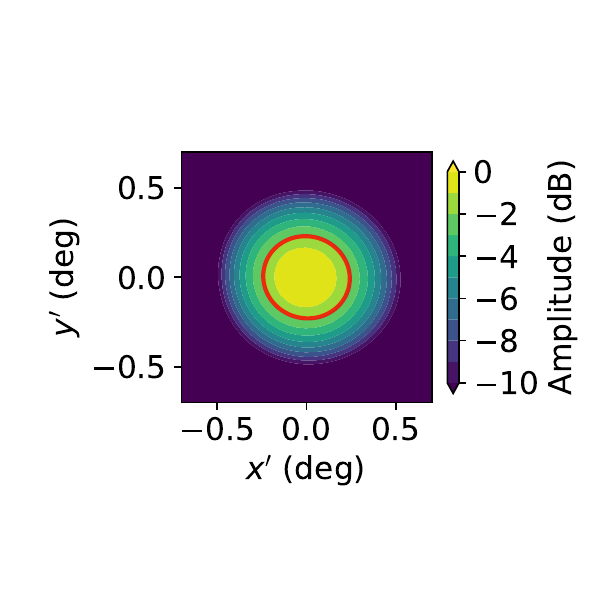}
\label{fig:beam2d_pixel5_18}}
\hspace{-0.2in} 
\subfloat[]{\includegraphics[width=2.4in]{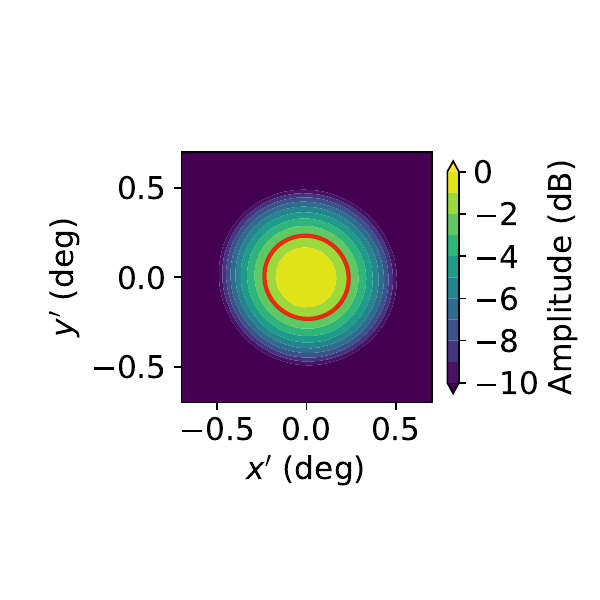}
\label{fig:beam2d_pixel5_wide}}
\caption{Two-dimensional beam patterns of pixel ID 5 at 145 GHz for different baffle aperture angles. The coordinates $(x', y')$ represent the angular offset from the beam center in azimuth and elevation directions. The red contours show the ellipse of the fitted 2D Gaussian function defined by $(x' \cos\delta + y' \sin\delta)^2/\sigma_1^2 + (-x' \sin\delta + y' \cos\delta)^2/\sigma_2^2 = 1$, which represents 
the beam shape.
(a) $\Theta = 15^\circ$ (ellipticity 0.13).
(b) $\Theta = 18^\circ$ (ellipticity 0.03).
(c) Wide baffle $\Theta = 27^\circ$ (ellipticity 0.02).
\label{fig:beam2d_pixel5}}
\end{figure*}
The GRASP simulations were performed for various $\Theta$. 
Fig.~\ref{fig:crosec_spill_0} shows the cross sections of the beam patterns on the line passing through the peak position of the spillover component when $\Theta=27^{\circ}$ and the center of the map that is $\theta = 0$ for the center pixel of ID 0.
The beam pattern for this pixel is calculated at the frequency of 220 GHz.
There is a prominent excess power below $-15$ degrees in the beam pattern for $\Theta=27^{\circ}$.
It is attributed to the spillover component. 
Fig.~\ref{fig:crosec_spill_0} also shows that the amplitude of the spillover components decreases as the baffle aperture angle $\Theta$ reduces.
For the baffles with $\Theta < 18^\circ$, the amplitude of the spillover component is suppressed by about $- 30$ dB from  that when the baffle aperture is set to  $\Theta=27^\circ$.

Fig.~\ref{fig:crosec_spill_5} shows the cross section of the beam patterns cut by the same way as Fig.~\ref{fig:crosec_spill_0} but for pixel 5. 
The beam was calculated at the frequency of 145 GHz. 
The spillover contribution can be seen below $\theta=-15^\circ$ for the baffle with $\Theta=27^\circ$, although the contrast relative to the amplitude of the side lobe of the main beam is not so prominent.
It does not change when $\Theta$ is decreased to $20^\circ$.
When the baffle with $\Theta=18^\circ$ is adopted, it is suppressed by about $-10$ dB from that. 
For both central and edge pixels, the spillover contamination is almost perfectly suppressed when we set $\Theta=15^\circ$. 

Fig.~\ref{fig:edge_near} shows that the near field main beam is diffracted by the baffle significantly when $\Theta=15^\circ$. 
It indicates that the far field main beam pattern could be significantly disturbed when the aperture angle of the baffle is set to be as narrow as $\Theta=15^\circ$. 
To quantify how the quality and the full width at half maximum~(FWHM)
of the main beam is degraded by narrowing the aperture angle of the baffle, 
$\Theta$ dependence of the ellipticity of the simulated main beam was evaluated. 
The main beam patterns were fitted with a 2D Gaussian function:
\begin{multline}
    G(x, y) = \exp\left[-\frac{(x^\prime \cos\delta + y^\prime \sin\delta)^2}{2\sigma_1^2} \right.\\
    \left. -\frac{(- x^\prime \sin\delta + y^\prime \cos\delta)^2}{2\sigma_2^2}\right]
\end{multline}
where $x$ and $y$ coordinates are set to azimuth and elevation directions, respectively, 
$(x_c, y_c)$ represents the position of maximum beam intensity, and $\delta$ is the angle of the major axis relative to $x$ coordinate.
The primed coordinates are defined as $x^\prime = x - x_c$ and $y^\prime = y - y_c$.
The ellipticities of the main beams are defined by $|\sigma_1 - \sigma_2|/(\sigma_1 + \sigma_2)$, which ranges from 0~(perfectly circular) to 1~($\sigma_1 >> \sigma_2$, or vice versa).
Fig.~\ref{fig:beam2d_pixel5} shows representative examples of the two-dimensional beam patterns for pixel ID 5 under three different baffle configurations. The fitted ellipses clearly visualize the degradation of beam circularity as the baffle aperture narrows from $27^\circ$ to $15^\circ$, with the ellipticity increasing from 0.02 to 0.13.
Table~\ref{tab:ellipticity} summarizes the ellipticity values for selected detector pixels at aperture angles of $15^\circ$, $18^\circ$, and $27^\circ$~(the wide baffle) to demonstrate the trend across different aperture configurations.
The main beam shapes were maintained with a constant FWHM and ellipticity for the central pixel~(pixel ID 0), regardless of the angles of the baffle aperture.
The main beam shapes deviate from a perfect circle for all pixels even when the wide baffle is installed.
With the wide baffle, the deviation from a circle is the worst for Pixel ID 1, which has an ellipticity of 0.03.
When the baffle with $\Theta=18^\circ$ is installed, pixel IDs 1 and 4 show the worst ellipticity of 0.04. 
When the baffle with $\Theta=15^\circ$ is installed, pixel ID 2 shows the worst ellipticity of 0.18.
We have checked that the FWHM of main beams does not change significantly for all analyzed pixels as far as $\Theta \ge 18^\circ$.

Based on these comprehensive analyses of beam quantification, thermal loading, and far field beam patterns, we concluded that the optimal aperture angle of the baffle is $\Theta = 18^{\circ}$. At this angle, the spillover component can be suppressed while maintaining the quality of the main beam and keeping the thermal loading from the baffle below 15~\% of the atmospheric contribution.
\begin{table}[t!]
\caption{Ellipticity of main beams for different baffle aperture angles.\label{tab:ellipticity}}
\centering
\begin{tabular}{|c|c|c|c|}
\hline
Pixel ID & $\Theta=15^\circ$ & $\Theta=18^\circ$ &  The wide baffle\\
\hline
0 & 0.01 & 0.01 &  0.02 \\
1 & 0.10 & 0.04 & 0.03 \\
2 & 0.18 & 0.01 & 0.02 \\
3 & 0.07 & 0.01 & 0.02 \\
4 & 0.11 & 0.04 & 0.02 \\
5 & 0.13 & 0.03 & 0.02 \\
6 & 0.15 & 0.03 & 0.02 \\
7 & 0.15 & 0.03 & 0.02 \\
8 & 0.13 & 0.03 & 0.02 \\
9 & 0.02 & 0.01 & 0.01 \\
\hline
\end{tabular}
\end{table}

\section{Optimized Baffle Design}
\begin{figure}[!t]
\centering
\subfloat[]{\includegraphics[width=0.45\linewidth]{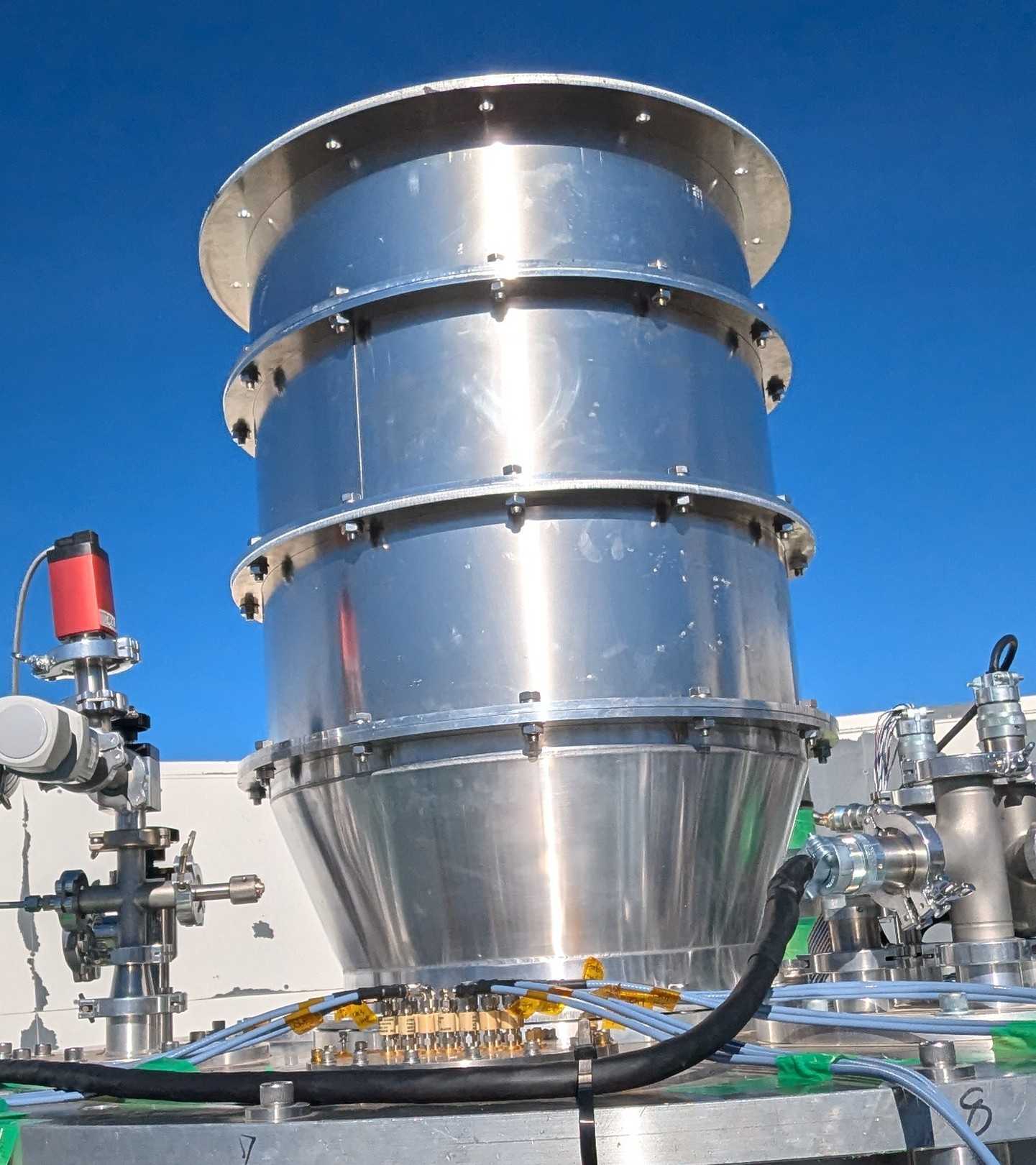}
\label{fig:NewBaffle_out}}
\hfill
\subfloat[]{\includegraphics[width=0.45\linewidth]{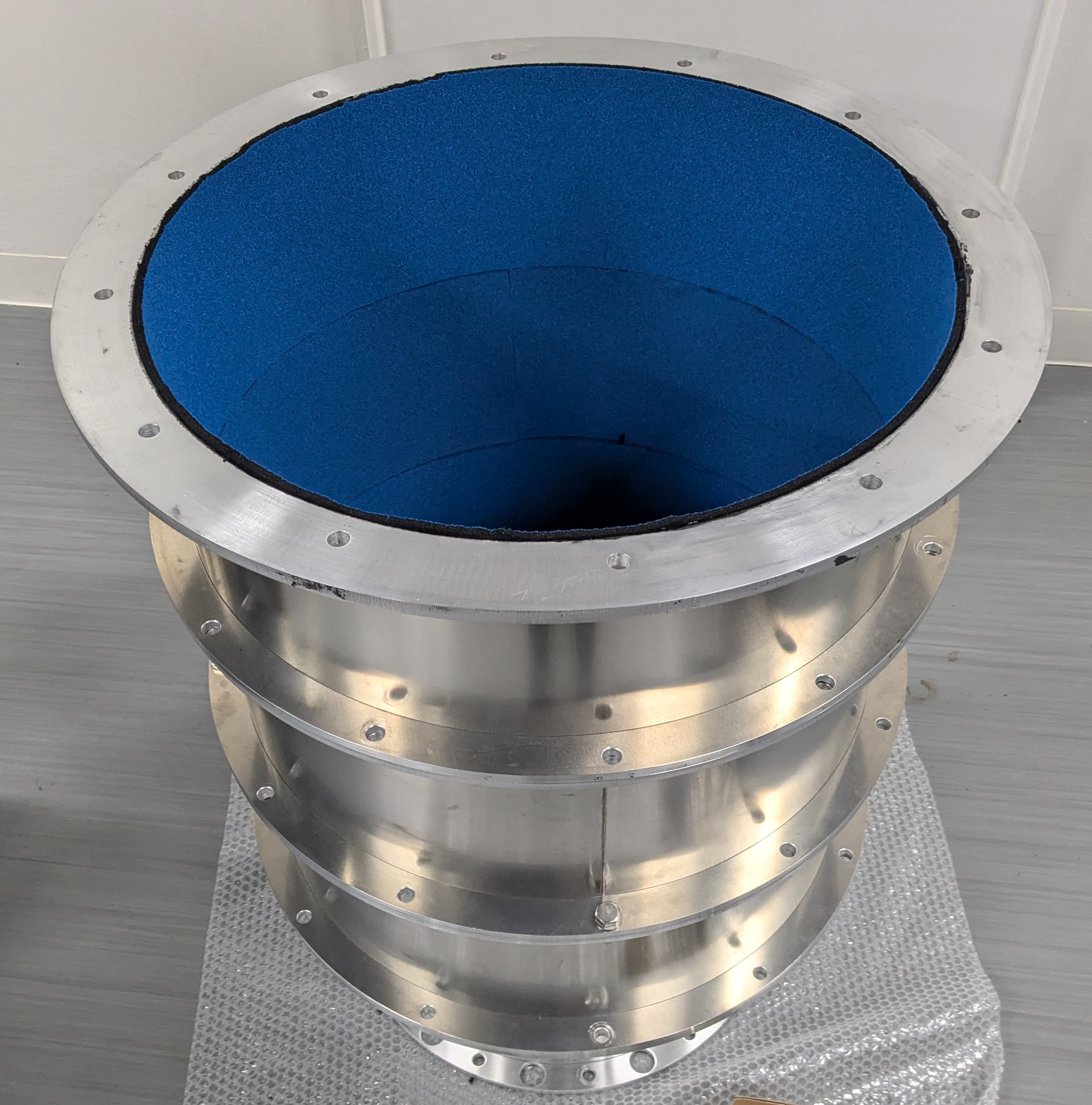}
\label{fig:NewBaffle_in}}
\caption{A photograph of the optimized baffle.
(a) External view showing the multi-stage cylindrical structure.
(b) Internal view showing the radio-absorptive material (ECCOSORB AN-72) covering the inner wall. The absorber is bonded to the aluminum surfaces using Stycast 2850FTJ adhesive.}
\label{fig:NewBaffle}
\end{figure}

Based on the results presented in the previous section, 
the optimized baffle was designed with an aperture angle of approximately $18^\circ$ to achieve maximum performance.
Fig.~\ref{fig:NewBaffle} shows a photograph of the optimized baffle. 

Our design adopted a multi-stage cylindrical structure rather than the conical geometry adopted in the wide baffle, as shown in Fig.~\ref{fig:NewBaffle_out}.
The baffle is constructed of multiple pieces connected by flanges to facilitate installation.
A cylindrical structure was selected to facilitate the installation of absorbers on the baffle wall~(see Section~\ref{subsec:thermalload}).
The internal view of the baffle is shown in Fig.~\ref{fig:NewBaffle_in}.
Since the thermal loading from the baffle wall only depends on the area of the baffle wall perpendicular to the near-field beam, this structural change does not increase the thermal loading compared to a conical structure baffle with the same aperture diameter. 

The optimized baffle geometry features are as follows. 
The baffle length, $l$, was set to be $l=82$~cm so as not to interfere with the dome.
The diameter at the top of the baffle was 54~cm.  
After lining with the absorber with a thickness of 0.6 cm, the diameter at the top of the baffle aperture, $D$,was reduced to $\sim$52.8~cm. 
Therefore, the aperture angle is $\Theta = \arctan\left(\frac{D/2}{l}\right) = \arctan\left(\frac{52.8/2}{82}\right) \sim 17.8^{\circ}$.

\section{Performance Evaluation}
\subsection{Stray Light Reductions}
\begin{figure}[!t]
\centering
\includegraphics[width=2.6in]{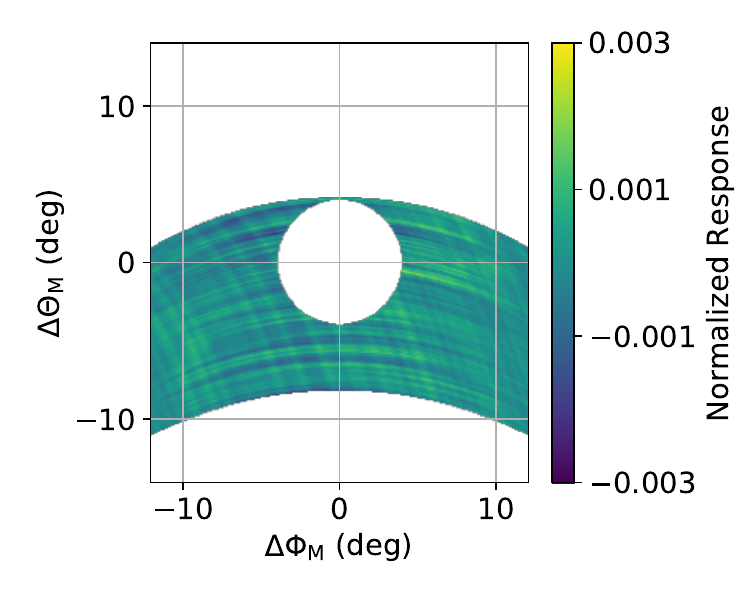}
\caption{Observed map of the Moon obtained by the pixel ID 9 when the optimized baffle is mounted on the GroundBIRD. The map is produced with the same procedures as Fig.~\ref{fig:moon_stray_original}.}
\label{fig:moon_stray_optimized}
\end{figure}

Fig.~\ref{fig:moon_stray_optimized} shows a Moon-centered map obtained with detector pixel ID 9 using the optimized baffle.
The data were acquired during observations on May 4, 2025, from 20:49 to 21:49 UTC.
The telescope rotation speed was 5~RPM. 
The average value of the PWV was around 2~mm.
The ghost image of the Moon, caused by the spillover component, was clearly suppressed compared to Fig.~\ref{fig:moon_stray_original}. 
The maximum noise amplitude is around $-25$ dB, which is similar to that shown in Fig.~\ref{fig:moon_stray_original}.
We confirmed that the ghost images of the Moon were suppressed for the other 4 pixels mounted on the same chip as pixel ID 9, as expected from the simulation.

\subsection{Performance of the Thermal Loading}
\begin{figure}[!t]
\centering
\includegraphics[width=3.in]{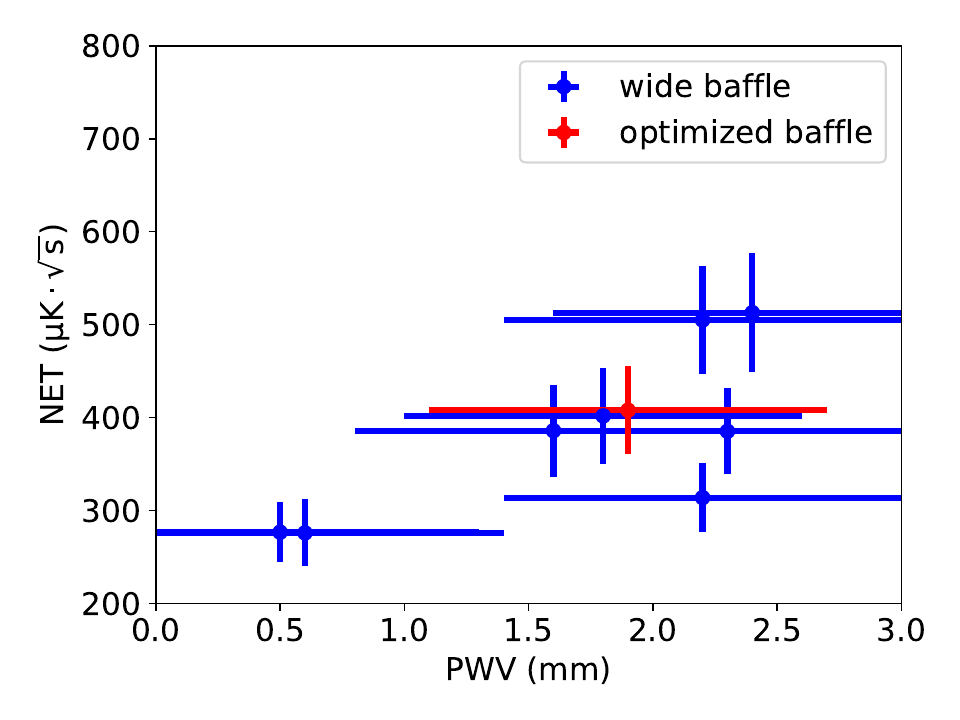}
\caption{The PWV dependence of the NET for the detector ID 9. The NET is given in units of Rayleigh-Jeans temperature. Results obtained for various PWVs with the wide baffle are shown by blue points. A result obtained with the optimized baffle is shown by a red point.}
\label{fig:NET}
\end{figure}

We evaluated the optical loading budget by comparing the NETs of the two different baffle configurations~(optimized and wide baffles).
The methods of extracting NET using Moon observations are given in our previous study\cite{GBMKID}.
The error bars for NET are derived from the root mean square of the white noise component in the power spectral density~(PSD) over the frequency range of 100 -- 500 Hz.
The error bars for PWV are derived from PWV measurements at Teide Observatory \cite{gaulli, CastroAlmazan2016}.
Fig.~\ref{fig:NET} shows the NET for various PWVs.
There is a PWV dependence of NETs measured with the wide baffle. It confirms that the dominant noise source is the atmospheric emission in the case of the wide baffle. 
The NET obtained with optimized baffle is consistent with that obtained with the wide baffle when $\mathrm{PWV}=1.9$ mm. This confirms that the noise loading due to thermal emission from the absorbers attached to the baffle is negligibly small compared with the noise due to atmospheric emission at $\mathrm{PWV}=1.9$ mm.
According to the GroundBIRD forecast paper\cite{lee2021forecast}, the scientific goal of the GroundBIRD is achieved by three-year observations under atmospheric conditions of $\mathrm{PWV}=4$ mm.
The average PWV at the Teide Observatory from 2023 July to 2024 August was 4.3 mm\cite{jo2025systematics}.  
We concluded that the thermal loading due to the optimization of the baffle design has a negligible impact on achieving the scientific goal of the GroundBIRD.
This was confirmed for other active pixels mounted on the same chip as ID 9.

\section{Discussion and Conclusion}
We have presented the optimization procedures of the baffle mounted on the GroundBIRD telescope, which employs a dual-mirror reflector telescope installed in the cryostat. 
The key strategies were to reduce stray light contamination as much as possible while maintaining the quality of the main beam and keeping the thermal loading from the baffle significantly below the atmospheric thermal loading.
Using quasi-optics simulations, we optimized the baffle’s aperture angle to suppress stray light without degrading the main beam quality. 
We confirmed through Moon observations that the optimized baffle design works to eliminate stray light contamination as expected.
Additionally, there was no degradation in NET, indicating minimal thermal impact. Although simulation and observational data suggest that the optimized baffle does not affect the main beam, further tests with a far-field point source would be beneficial. However, due to the absence of a bright celestial point source for GroundBIRD, alternative methods such as drone-mounted artificial sources may be necessary\cite{dunner2021drone, coppi2022protocalc}. This work not only improves the current performance of the GroundBIRD telescope but also offers practical guidance for future CMB experiments.

\section*{Acknowledgment}
This work was supported by the GroundBIRD collaboration.
This work was supported by MEXT KAKENHI Grant Number JP18H05539 and JSPS KAKENHI Grant Numbers JP15H05743, JP20K20927, JP20KK0065, JP21H04485, JP21K03585, JP22H04913, and JP24H00224, JSPS Bilateral Program Numbers JPJSBP120219943 and JPJSBP120239919, and also supported by JSPS Core-to-Core Program JPJSCCA20200003.
M.T. also acknowledges support from Graduate Program on Physics for the Universe (GP-PU), Tohoku University, and JST SPRING, Grant Number JPMJSP2114.
We thank Victor González Escalera, \'{A}ngeles P\'{e}rez de Taoro (the Instituto de Astrof\'{i}sica de Canarias), and the staff of Teide Observatory for supporting the maintenance and operation of GroundBIRD.
M.T. gratefully acknowledges the Machine shop, School of Science, Tohoku University and G. Nakamura for their valuable consultation and technical assistance in the baffle fabrication process.
We thank anonymous referees for providing constructive comments. These were very helpful in revising our paper.

\end{document}